\documentclass[nobib,nohyper]{tufte-handout}
\usepackage[LGR,T1]{fontenc}
\usepackage[utf8]{inputenc}
\newcommand{\textgreek}[1]{\begingroup\fontencoding{LGR}\selectfont#1\endgroup}

\usepackage{amsthm}
\usepackage{amssymb}
\usepackage{amsmath}

\usepackage{tikz-cd}

\theoremstyle{plain}
\newtheorem{theorem}{Theorem}

\theoremstyle{definition}
\newtheorem{definition}[theorem]{Definition}

\newtheorem{principle}[theorem]{Principle}

\newcommand{\N}{\mathbb N}

\newcommand{\B}[1]{\textbf{#1}}

\title[The Algebraic view of Computation]{The Algebraic View of Computation}
\date{}
\author[e-n@]{Attila Egri-Nagy}

\usepackage{graphicx} 
  \setkeys{Gin}{width=\linewidth,totalheight=\textheight,keepaspectratio}
  \graphicspath{{graphics/}} 
\usepackage{amsmath}  
\usepackage{booktabs} 
\usepackage{units}    
\usepackage{multicol} 
\usepackage{lipsum}   
\usepackage{fancyvrb} 
  \fvset{fontsize=\normalsize}

\newcommand{\cT}{\mathcal T}

\newcommand{\Z}{\mathbb{Z}}

\begin{document}

\maketitle

\begin{abstract}
We argue that computation is an abstract algebraic concept, and a computer is a result of a morphism (a structure preserving map) from a finite universal semigroup.
\end{abstract}


The steam engine replaced muscle power. It did not just make life easier, but a whole bunch of impossible things became possible.
Curiously, it was invented before we understood how it worked.
Then, trying to make it more efficient led to thermodynamics and indirectly to a deeper understanding of the physical world.
Similarly, computers replace brain power, but we still do not have a full comprehension of computation.
Trying to make computation more efficient and to find its limits is taking us to a deeper understanding of not just computer science but of other branches of science (e.g.~biology, physics, mathematics).
Just as physics advanced by focusing on the very small (particles) and on the very large (universe), studying computers should also focus on the basic building blocks (finite state computations) and on the large abstract structures (hierarchical (de)compositions).
Another parallel with physics is that the underlying theory of computation is mathematical.
The following theses summarize the key points of the algebraic view of computation:
\begin{enumerate}

\item \textbf{Computation has an abstract algebraic structure.} Semigroups (sets with associative binary operation) are natural generalizations of models of computation (Turing machines, $\lambda$-calculus, finite state automata, etc.).
\item \textbf{Algebraic structure-preserving maps are fundamental for any theory of computers.}
Being a computer is defined as being able to emulate/implement  other
computers, i.e.~being a homomorphic/isomorphic image.
The rule of keeping the 'same shape' of computation applies without exception,  making programability and interactivity possible.
\item \textbf{Interpretations are more general functions than
    implementations.} An arbitrary function without morphic properties can define semantic content for a computation, or we  can map a single trace of execution only. These cannot guarantee the result of the map being a computer.
\item \textbf{Computers are finite.} Finiteness renders decision problems trivial to solve, but computability with limited resources is a fundamental engineering problem that still requires mathematical research.

\item \textbf{Computers are universal.}
In a sense every dynamical system computes something, namely its future states, but in order to be a computer we require that it should be able to compute everything else within its finite limits.
\item \textbf{Hierarchy is an organizing principle of computation.}
  Artificial computing systems tend to have one-way (control) information flow for modularity. Natural systems with feedback loops also admit hierarchical models.

\end{enumerate}

Here we reflect on the worldview, on the tacitly assumed ontological stance of a computational mathematician and software engineer, who is chiefly concerned with extending mathematical knowledge by enumerating finite structures.
As such, we will mainly focus on classical digital computation.
However, semigroup theory is abstract enough to accommodate other kinds of computations.
We will use only minimal mathematical formalism here; for technical details see \cite{candar16,egri2017finite}.

With another physical metaphor, we can say that here we present the `particle physics' of computation, the ultimate underlying mathematics of computers.
Therefore, many aspects of computation (e.g.~engineering practices, its social impact, etc.) will not be discussed here.

First we generalize traditional models of computation to composition tables, that describe semigroups.
We show how these abstract semigroups can cover the wide spectrum of computational phenomena.
Then we argue that homomorphisms, structure preserving maps between semigroups, together with finite universality give the important concepts for defining computers.
We mention how this fundamental theory can give rise to higher level of computational structures and touch upon several philosophical and more open ended considerations.

\section{Semigroup -- composition table of computations}

The Turing machine \cite{Turing1936,bernhardt2016turing} is a formalization of what a human calculator or a mathematician would do when solving problems using pencil and paper. As a formalization, it abstracts away unnecessary details. For instance,  the fact that we write symbols line by line on a piece of paper, the two-dimensional nature of the sheet can be replaced with a one-dimensional tape.
Also, for symbolic calculations, the possibly infinite spectrum of moods and thoughts of a calculating person can be substituted with a finite set of distinguishable states.
A more peculiar abstraction is the removal of the limits of human memory. This actually introduces a new feature: infinity. This is coming from the purpose of the model (to study decidability problems in logic) rather than the properties of a human calculator.
Once we limit the tape or the number of calculation steps to a finite
number, the halting problem ceases to be undecidable.
In practice, we can do formal verification of programs.
Therefore, to better match existing computers, we assume that memory capacity is finite.
Such a restricted Turing machine is a finite state automaton (FSA) \cite{minsky1967computation}.

Now the FSA still has a lot that can be abstracted away. The initial and accepting states are for recognizing languages. The theory of formal languages is a special application of FSA.
The output of the automaton can be defined as an additional function of the internal states, so we do not have to define output alphabet.
What remains is a set of states, a set of input symbols and a state transition function. An elementary event of computation is that the automaton is in a state, then it receives an input and based on that it changes its state.
We will show that the distinction between input symbols and states can also be abstracted away.
It is important to note that FSA with a single input (e.g.~clock-tick, the passage of time) are only a tiny subset of possible computations. Their algebraic structure is fairly simple. Metaphorically speaking they are like batch processing versus interactivity.

\emph{What is then the most general thing we can say about computation?} It certainly involves change.
\marginnote{``To compute is to execute an algorithm.'' \cite{Copeland1996}}
Turning input into output by executing an algorithm and going through many steps while doing so can be described as a sequence of state transitions.
We can use the fundamental trick of algebra (writing letters to denote a whole range of possibilities instead of a single value) and describe an elementary  event of computation by the equation
$$xy=z.$$
Abstractly, we say that $x$ is combined with $y$ results in $z$.
The composition is simply denoted by writing the events one after the other.
One interpretation is that event $x$ happens, then it is followed by event $y$, and the overall effect of these two events combined is the event $z$.
Or, $x$ can be some input data and $y$ a function (in the more usual notation it would be $y(x)$).
Or, the same idea with different terminology, $x$ is a state and $y$ is a state-transition operator. This is the answer for the \emph{How to compute?} question.
If we focus on the question \emph{What to compute?}, then we are interested only in getting the output from some input. \marginnote{``Abstract computers (such as finite automata and Turing machines) are essentially function-composition schemes.'' \cite{toffoli1980reversible}}
In this sense, computation is function evaluation, as in the mathematical notion of a function.
We have a set of inputs, the \emph{domain} of the function,  and a set of outputs, the \emph{codomain}.
\marginnote{``Intuitively, a computing machine is any physical system whose dynamical evolution takes it from one
of a set of `input' states to one of a set of `output' states.'' \cite{Deutsch85}}
We expect to get an output for all valid inputs, and for a given input we want to have the same output whenever we evaluate the function.
In practical computing we often have `functions' that seem to violate these rules, so we distinguish between pure functions.
A function call with side-effect (e.g.~printing on screen) is not a mathematical function.
This depends on how we define the limits of the system. If we put the current state of the screen into the domain of the function, then it becomes a pure function.
For a function returning a random number, in classical computation it is a pseudo-random number, so if we include the seed as another argument for the function, then again we have a pure function.

\begin{marginfigure}
  \begin{tabular}{r|ccc}
flip-flop &$r$&$s_0$&$s_1$\\
\hline
$r$&$r$&$s_0$&$s_1$\\
$s_0$&$s_0$&$s_0$&$s_1$\\
$s_1$&$s_1$&$s_0$&$s_1$
\end{tabular}
\vskip1ex

\begin{tabular}{r|ccc}
$\Z_3$ &$+0$&$+1$&$+2$\\
\hline
$+0$&$+0$&$+1$&$+2$\\
$+1$&$+1$&$+2$&$+0$\\
 $+2$&$+2$&$+0$&$+1$
\end{tabular}
\caption{Composition (multiplication) tables of semigroups (computational structures). The flip-flop is the semigroup of a 1-bit memory device ($r$ -- reading the content, $s_0,s_1$ -- storing bit $0$ and bit $1$ destructively).
  $\Z_3$ is a modulo-3 counter, i.e.~an odometer with only three possible values.}
\label{fig:multabs}
\end{marginfigure}
A single composition, when we put together $x$ and $y$ (in this order) yielding $z$, is the elementary unit of computation. These are like elementary particles in physics. In order to get something more interesting we have to combine them into atoms. The atoms of computation will be certain tables of these elementary compositions, where two conditions are satisfied.
\begin{enumerate}
  \item The composition has to be \emph{associative}
$$x(yz)=(xy)z$$
meaning that a sequence of compositions $xyz$ is well-defined.
\marginnote{``A computation is a process that obeys finitely describable rules.'' \cite{rucker2006lifebox}}
\item The table also has to be \emph{self-contained}, meaning that the result of any composition should also be included in the table. Given a finite set of $n$ elements, the $n\times n$ square table will encode the result of combining any two elements of the set.
\end{enumerate}
  \marginnote{``Numbers measure size, groups measure symmetry.'' \cite{armstrong1988groups} -- and semigroups measure computation.}
The underlying algebraic structure is called the \emph{semigroup} (a
set with an associative binary operation, \cite{Howie95}), and the composition is often called multiplication (due to its traditional algebraic origin), or the Cayley-table (Fig.~\ref{fig:multabs}).
Continuing the physical metaphor, not all composition tables are atoms, as some tables are built by using simpler tables (as we will discuss later).

Talking about state transitions, we still do not say anything concrete about the states. If state changes along a continuum, then we talk about analog computing. If state is a discrete configuration then we have classical computing. In case we have a vector of amplitudes, then we have quantum computing.
Also, $xy$ in itself is a sequential composition, but $y$ can be a parallel operation. We will see that concurrency and parallelism are more specific details of computations.

\emph{Is the distinction between states and events fundamental?}
The algebraic thinking guides the abstraction. The number $4$ can be identified with the operation $+4$, relative to $0$, so it is both a state and an operation.
\begin{principle}[State-event abstraction]
  We can identify an event with its resulting state: state $x$ is where we end up when event $x$ happens, relative to a ground state. The ground state in turn corresponds to a neutral event, that does not change any state.
\end{principle}

\subsection{Accretion of structure}
The classical (non-interactive) computation admits another characterization: it is generating structures from partial descriptions.
The archetypical example is a sudoku puzzle with a unique solution \cite{sciam-sudoku}, where the accretion of the structure is visual: more numbers are put into the table.
Even when we only keep the final result as a single data item, we still generate intermediate data (structure).
More general examples are graph search algorithms, where the graph is actually created during the search.
Logical inference also fits this pattern: the premises determine the conclusions through intermediate expressions.
The existing entries (input) implicitly determine the whole table (output) but we have to execute an algorithm to find those entries. As a seed determines how a crystal grows, the input structure determines the whole.

This idea has a clear algebraic description: the set of \emph{generators}. These are elements of a semigroup whose combinations can generate the whole table. In order to calculate the compositions of generators they have to have some representation. For instance, transformations of a finite set with $n$ elements. A transposition, a full cycle, and an elementary collapsing can generate all possible $n^n$ transformations.
For a more general computation, the executable program and the input data together serve as a generating set, or the primitives of a programming language can take that role \cite{tslang2010}.

\subsection{Timeless computation?}

The accretion of structure view of computation has an interesting interplay with time.
When executing an algorithm that generates a full structure from a partial description, in a sense the structure is already there. When a computational experiment is set up and the programmer hits the ENTER key, all that separates her from knowing the answer is time.
Time is crucial for computation.
Much of computer science and software engineering is about solving problems faster.
Computational complexity classifies algorithms by their space and time requirements.
Often space can be exchanged for time and the limit of this process is the lookup table, the precomputed result. Information is frozen computation.
Taking the abstraction process to its extreme, we can replace the two-dimensional composition table with a one-dimensional lookup table, with keys as pairs $(x,y)$ and values $xy$.
At the very bottom computation is just association, keys to values. This explains why arrays and hashtables are important data structures in programming.
The composition table is just an unchanging array, thus all computations of a computing device have a timeless interpretation as well.
Here we do not want to go into the several interpretations of time (for the two extremes see \cite{smolin2013time, barbour2001end}), just to emphasize that computation is orthogonal to the problem of time. We can talk about static \emph{computational structures}, composition tables, and we can also talk about computational processes, sequences of events tracing a path in the composition table.

\section{Homomorphism -- the algebraic notion of implementation}
\marginnote{``In enabling mechanism to combine together general symbols in successions of unlimited variety and extent, a uniting link is established between the operations of matter and the abstract mental processes of the most abstract branch of mathematical science.'' \cite{AdaNotes}}
Homomorphism is a simple concept, but its significance can be hidden in the algebraic formalism.
The etymology of the word conveys the underlying intuitive idea: the ancient Greek \textgreek{ὁμός} (homos) means `same' and \textgreek{μορφή} (morphe) means `form' or `shape'. Thus, homomorphism is a relation between two objects when they have the same shape.
The abstract shape is not limited to static structures, thus we can talk about homomorphisms between dynamical systems, i.e.~finding correspondences between states of two different systems and for their state transition operations as well. Change in one system is mimicked by the change in another.
Homomorphism is a knowledge extension tool: we can apply knowledge about one system to another. It is a  way to predict outcomes of events in one dynamical system based on what we know about what happens in another one, given that a homomorphic relationship has been established. It is also a general trick for problem solving widely used in mathematics. If obtaining a solution is not feasible in one problem domain, then by transferring the problem to another domain we can use easier operations, given that we can move between the domains with structure preserving maps.

What does it mean to be in a homomorphic relationship for computational structures?
 \marginnote{``A physical system implements a given computation when the causal structure of
   the physical system mirrors the formal structure of the computation.'' \cite{Chalmers1994Implementation}}
 Using the composition table definition we can now define their structure preserving maps. If in a system $S$ event $x$ combined with event $y$ yields the event $z=xy$, then by a homomorphism $\varphi:S\rightarrow T$, then in another system $T$ the outcome of $\varphi(x)$ combined with $\varphi(y)$ is bound to be $\varphi(z)=\varphi(xy)$, so the following equation holds
$$\varphi(xy)=\varphi(x)\varphi(y).$$
On the left hand side, composition happens in $S$, while on the right hand side composition is done in $T$ (for example Fig.~\ref{fig:embedding}).
What is the typical usage of the homomorphism?
Let's say I want to compute $xy$, where $x$ can be some input data and $y$ a function.
But I cannot just apply the function, because it would be impossible to do it in my head or it would take a long time to do it on sheets of  paper with a pen. But I have some physical system $T$, whose internal dynamics is homomorphic how the function works. So I represent $x$ in $T$ as $\varphi(x)$, and $y$ as $\varphi(y)$, then let the dynamics of $T$ carry out the combination of $\varphi(x)\varphi(y)$. By the homomorphism, that is the same as $\varphi(xy)$. At the end I need to find out how to map the result back to $xy$.
\begin{figure*}[t]
\begin{center}
\begin{tabular}{c|cc}
$\Z_2$ & 0 & 1 \\
\hline
0&0&1\\
1&1&0\\
\end{tabular}
$\hookrightarrow$
\begin{tabular}{c|cccc}
$\cT_2$ & 1 & \textbf{2} &\textbf{3} &4\\
\hline
1&1&1&4&4\\
\textbf{2}&1&\textbf{2}&\textbf{3}&4\\
\textbf{3}&1&\textbf{3}&\textbf{2}&4\\
4&1&4&1&4\\
\end{tabular}
\end{center}
\caption{The maps $0\mapsto 2$, $1 \mapsto 3$ define an isomorphism (embedding).}
\label{fig:embedding}
\end{figure*}

What makes homomorphism powerful is that it is \emph{systematic}. It works for all combinations not just a one-off correspondence. There is no way to opt out: the rule has to work not just for a single sequence but for all possible sequences of events, for the whole state-transition table.
Otherwise, one could fix an arbitrary long computation as a sequence of state transitions. Then, by a carefully chosen encoding, any physical system with enough states can execute the same sequence. But the same encoding will be unlikely to work for a different sequence of state transitions, thus it is not a homomorphism.
We argue, that the algebraic definition of homomorphism should form the base of the philosophical discussion, the starting point.
Without the precision of algebra it becomes possible to talk about computing rocks and walls, pails of water (for an overview of pancomputationalism see \cite{computation-physicalsystems}).

A distinguished class of homomorphisms are \emph{isomorphisms}, where the correspondence is one-to-one.
In other words, isomorphisms are strictly structure preserving, while homomorphisms can be structure forgetting down to the extreme of mapping everything to a single state and to the identity operation.
The technical details can be complicated  due to clustering states (surjective homomorphism) and by the need of turning around homomorphism we also consider homomorphic relations \cite{candar16}.
\marginnote{
  ``\ldots we need to
discover whether the laws of physics are prior to, in the sense of
constraining, the possibilities of computation, or whether the laws of
physics are themselves consequences of some deeper, simpler rules of step-by-step computation.'' \cite{PiInTheSky}

  At least in science-fiction, turning it around: mathematical truth (about abstract structures) depends on computers):
```A mathematical theorem,' she'd proclaimed, `only becomes true when a physical system tests it out: when the system's behaviour depends in some way on the theorem being \emph{true} or \emph{false}.  ''

``\ldots And if a mathematician could test those steps by manipulating a finite number of physical objects for a finite amount of time -- whether they were marks on paper, or neurotransmitters in his or her brain -- then all kinds of physical systems could, in theory, mimic the structure of the proof\ldots with or without any awareness of what it was they were proving'.''\cite{Luminous}
}

\subsection{Computers as Physical Systems}
The point of building a computer is that we want the computation done by a physical system on its own, just by supplying energy. So if a computational structure as a mathematical entity determines the rules of computation, then somehow the physical system should obey those rules.

\begin{definition}[vague]
Computers are physical systems that are homomorphic images of computational structures (semigroups).
\end{definition}
This first definition begs the question, how can a physical system be an image of a homomorphism, i.e.~a semigroup itself? How can we cross the boundary between the mathematical realm and the external reality?
First, there is an easy but hypothetical answer.
\marginnote{``Our external physical reality is a mathematical structure.'' \cite{TMU2008}}
According to the Mathematical Universe Hypothesis \cite{TMU2008, tegmark2014our}, all physical systems are mathematical structures, so we never actually leave the mathematical realm.

Secondly, the implementation relation can be turned around. Implementation and modelling are the two directions of the same isomorphic relation. If $T$ implements $S$, then $S$ is a computational model of $T$. Again, we stay in the mathematical realm, we just need to study mappings between semigroups.
Establishing and verifying a computational model of a physical system require scientific work (both theoretical and experimental) and engineering. The computational model of the physical system may not be complete. For instance, classical digital computation can be implemented without quantum mechanics.
\marginnote{``Computing processes are ultimately abstractions of physical processes: thus, a
comprehensive theory of computation must reflect in a stylized way aspects of
the underlying physical world.'' \cite{toffoli1982physics}
}
\marginnote{
  ``Our computers do no more than re-program a part of the
  universe to make it compute what we want it to compute.'' \cite{ComputableUniverseIntro}
}
\marginnote{
``A computer is an arrangement of some of
the material constituents of the Universe into a configuration whose
natural evolution in time according to the laws of Nature simulates
some mathematical process.'' \cite{PiInTheSky}
}
\begin{definition}
Computers are physical systems whose computational models are homomorphic images of semigroups.
\end{definition}
Computation is orthogonal to the problem whether mathematics is an approximation or a perfect description of physical reality.

For practical purposes, we are interested in implementations of computational structures that are in some sense universal.
\marginnote{``\ldots the universal computer can eventually do what any other computer can. In other words, given enough
time it is universal.'' \cite{deutsch1998fabric}}
In the finite case, for $n$ states, we require the physical system be able to implement $\cT_n$.

Every dynamical system computes something, at least its future states. \marginnote{``In a sense, nature has been continually computing the `next state' of the universe for billions of years; all we have to do -- and, actually, all we can do -- is `hitch a ride' on this huge ongoing computation , and try to discover which parts of it happen to go near to where we want.'' \cite{toffoli1982physics}}
The question is whether we can make a system compute something useful for us, how much useful computation can the system perform.
In the steam engine, every water molecule has some kinetic energy, but not all of them happen to bump into the piston. All others generate only waste heat by banging on the walls of the cylinder.
Similarly, it is not easy to find dynamical systems that do computation useful for us. We have to design and engineer those. A piece of rock is unchanging on the macro level, so it only implements the identity function. On the microscopic level it can be used for computing random numbers by measuring the vibration of its atoms. But it is not capable of universal computation.
Only carefully crafted pieces of rock, the silicon chips, can have this very special property.

Biological systems are also good candidates for hosting computation, since they're already doing some information processing. However, it is radically different from digital computation. The computation in digital computers is like toppling dominoes, a single sequence of chain reactions of bit-flips. Biological computation is done in a massively parallel way (e.g.~all over in a cell), more in a statistical mode.

Alternatively, we can redefine the computational work we want to do.
If it is a continuous mathematical problem, then it is easy to find a physical system that is capable of the corresponding analogue computation.

\subsection{Interpretations}

Computational \emph{implementation} is a homomorphism, while an arbitrary function with no homomorphic properties is an \emph{interpretation}, we can just take a computational structure and assign some meaning to its elements, the semantic content. This map is not necessarily structure preserving. For instance, reversible system can carry out irreversible computation by a carefully chosen output encoding \cite{toffoli1982physics,egri2017finite}.
This in turn demonstrates that today's computers are not based on the reversible laws of physics. Computers dissipate heat. We implement semigroups by thermodynamical processes. It is an open problem, whether we can implement group computation with reversible transformations (and hook on a non-homomorphic function to extract semantic content).
In essence, the problem of reversible computation implementing programs with memory erasure is the same as trying to explain the arrow of time arising from the symmetrical laws of physics.

Interpretations look more powerful since they can bypass limitations, like implementing many-to-one functions using 1-to-one mappings.
However, since they are not necessarily structure preserving, the knowledge transfer is just one way.
If we ask a new question, then we have to devise a new encoding for the possible solutions.

\section{High-level structure: hierarchies}

Composition and lookup tables are the ``ultimate reality'' of computation, but  they are not adequate descriptions of practical computing.
The low-level process in a digital computer, the systematic bit flips in a vast array of memory, is not very meaningful.
\marginnote{``Computers are built up in a hierarchy of parts, with each part repeated many times over.'' \cite{hillis1998pattern}}
The usefulness of a computation is expressed at several hierarchical layers above (e.g.~computer architecture, operating system, end user applications).
Parallel computation, and interactive processes, while can be described as a gigantic composition table, they have more explanatory power when viewed by the rules their communication protocol.

First, an algebraic layer is needed for dealing with the multitude of computational events. In the form of equations, we need to express laws that are universal (e.g.~associativity), or specific to a particular computer.

Secondly, a semigroup is seldom just a flat structure, its elements may have different roles.
For example, if $xy=z$ but $yx=y$ (assuming $x\neq y\neq z$), then we say that $x$ has no effect on $y$ (leaves it fixed), while $y$ turns $x$ into $z$.
There is an asymmetric relationship between $x$ and $y$: $y$ can influence $x$ but not the other way around.
This unidirectional influence give rise to hierarchical structures.
It is actually better than that.
According to the Krohn-Rhodes theory \cite{wildbook} \emph{every} automaton can be emulated by a hierarchical combination of simpler automata.
This is true even for inherently non-hierarchical automata built with feedback loops between its components.
It is a surprsising result of algebraic automata theory that recurrent networks can be rolled out to one-way hierarchies.
These hierarchies can be thought as  easy-to-use cognitive tools for understanding complex systems \cite{NehanivCT97}.
They also give a framework for quantifying biological complexity \cite{BiologicalComplexity99}.

The simpler components are roughly of two kinds: groups and semigroups.
Now these can be viewed differently: the irreversible part is destructive memory storage, while the group is always reversible computation.
Groups are also associated with isomorphisms (due to the existence of uniqe inverses), therefore computation can also be viewed as pure data conversion.

\section{Wild considerations}

This theory of implementing computations is a description of what we have in today's computers.
It is not known whether the semigroup computation model could explain the mind, but there is not much left to abstract from.
Thus, the question  whether cognition is computational or not, is the same as the question whether mathematics is a perfect description of physical reality or just an approximation of it. If it is just an approximation, then there is a possibility that cognition resides in physical properties that are left out.

A recurring question in philosophical conversations is the possibility of the same physical system realizing two different minds simultaneously \cite{Shagrir2012}.
Let's say $n$ is the threshold for being a mind, so you need at least $n$ states for a computational structure to do so. Then supposedly there is more than one way to produce a mind with $n$ states, so the corresponding full transformation semigroup $\cT_n$ can have subsemigroups corresponding to several minds.
Then we need a physical system to implement $\cT_n$.
Now, it is a possibility to have different embeddings into the same system, therefore the algebra would allow the possibility of two minds coexisting in the same physical system.
This is how far mathematics can go in answering this question.

For scientific investigation these questions are still out of scope. Simpler ones do form a research program: like what is the minimum number of states to implement a self-referential system, or in general, what are the minimal implementations of certain functionalities, how many computational solutions are there for the same problem?

\section{Summary}

\marginnote{``\emph{Computation} A physical process that instantiates properties of some abstract entity.'' \cite{deutsch2011beginning}}

We showed that a generalization of models of computation to semigroups is flexible enough to cover the whole spectrum of computational phenomena.
Algebraic structure preserving maps spread `computerness' from one system to the other.
Starting from semigroups that have finite universality we can decide whether a given dynamical system is a computer or not.
Therefore, the algebraic viewpoint provides a solid base for further mathematical research and philosophical investigations of the computational phenomena.

\bibliography{comp}
\bibliographystyle{alpha}

\end{document}